\documentclass[aps,pre,amsmath,amssymb,twocolumn,groupedaddress,showpacs,eqsecnum]{revtex4}
\usepackage{graphicx}
\usepackage[utf8]{inputenc}
\usepackage[caption = false]{subfig}
\usepackage{float}

\usepackage{epsf}
\usepackage[toc]{appendix}
\newcommand{\be}{\begin{equation}}
\newcommand{\ee}{\end{equation}}
\newcommand{\ba}{\begin{eqnarray}}
\newcommand{\ea}{\end{eqnarray}}
\newcommand{\baa}{\begin{eqnarray}}
\newcommand{\eaa}{\end{eqnarray}}
\newcommand{\ed}{\end{document}}

\renewcommand{\baselinestretch}{1.2}
\date{\today}

\begin{document}

\title{Fast forward approach to stochastic heat engine}
\author{Katsuhiro Nakamura$^{1,2}$,  Jasur Matrasulov$^{1}$,  and Yuki Izumida$^{3}$ }
\affiliation{$^{1}$Faculty of Physics, National University of Uzbekistan, Vuzgorodok, Tashkent 100174, Uzbekistan\\
$^{2}$Department of Applied Physics, Osaka City University, Sumiyoshi-ku, Osaka 558-8585, Japan \\
$^{3}$Department of Complexity Science and Engineering, Graduate School of Frontier Sciences, The University of Tokyo, Kashiwa, Chiba 277-8561, Japan
}

\begin{abstract} The fast-forward (FF) scheme proposed by Masuda and Nakamura (\textit{Proc. R. Soc. A}  \textbf{466}, 1135  (2010)) in the context of conservative quantum dynamics can reproduce a quasi-static dynamics in an arbitrarily short time. We apply the FF scheme to the  classical stochastic Carnot-like heat engine which is driven by a Brownian particle coupled with a time-dependent harmonic potential and working between the high ($T_h$)- and low ($T_c$)-temperature heat reservoirs. Concentrating on  the underdamped case where momentum degree of freedom is  included, we find the explicit expressions for the FF protocols necessary to accelerate both the isothermal and thermally-adiabatic processes, and obtain the reversible and irreversible works.
The irreversible work is shown to consist of two terms with one proportional to and the other  inversely proportional to the friction coefficient. 
The optimal value of efficiency $\eta$ at the maximum power of this engine is found to be 
$\eta^*=\frac{1}{2} \left( 1+\frac{1}{2}\left(\frac{T_c}{T_h}\right)^{\frac{1}{2}} - \frac{5}{4}\frac{T_c}{T_h}  +O\left(\left(\frac{T_c}{T_h}\right)^{\frac{3}{2}}\right)\right)$ 
and $\eta^*= 1- \left(\frac{T_c}{T_h}\right)^{\frac{1}{2}}$, respectively in the cases of strong and weak dissipation.
The result is justified for a wide family of time scaling functions, making the FF protocols very flexible. 
We also revealed that the accelerated full cycle of the Carnot-like stochastic heat engine cannot be conceivable within the framework of the overdamped case, and the power and efficiency can be evaluated only when the momentum degree of freedom is taken into consideration. 
\end{abstract}
\maketitle

\section{Introduction}

Carnot's concept of heat engines belongs to a classical subject of thermodynamics. To achieve the highest efficiency, a heat engine needs to operate a reversible thermodynamic cycle which requires  a quasi-static process and results in a vanishing power. The power means the work per one-cycle time. The quasi-static thermodynamic cycle should be speeded up so as to produce a finite power of realistic heat engines. It is desirable to investigate how large the efficiency of a heat engine can be reached when the engine operates in the region of the maximum power. This issue has led to the birth of finite-time thermodynamics which has attracted much attention for many years. The most notable result in finite-time thermodynamics is the Curzon-Ahlborn (CA)'s efficiency, $\eta_{CA}\equiv1-\sqrt{T_c/T_h}$, which is the efficiency at maximum power for a macroscopic endo-reversible heat engine \cite{CA} operating between a cold bath at temperature $T_c$ and hot bath at temperature $T_h (> T_c)$.  CA noted the finite temperature difference between the heat bath and the working substance and took into consideration  the finite time needed for the heat transfer between them.

In contrast to the macroscopic heat engines considered in endo-reversible thermodynamics, thermal fluctuations play a crucial role in nano-scale systems, where dynamics cannot be described on a deterministic (macroscopic) level.  Sekimoto's stochastic energetics \cite{Sekimoto1,Sekimoto2,seki-hond} is a key to thermodynamic description of Langevin systems driven far from equilibrium, which can define thermodynamic quantities on a single stochastic trajectory \cite{Sekimoto2,Schmiedl,Seifert2} and  yield the ensemble quantities after averaging.

In the context of nano-scale motors, Brownian heat engines have received a wide attention, which mimic a simple system of a stochastic heat engine whose degrees of freedom are subject to a time-dependent potential and working between hot and cold heat baths. The efficiency of the engines of this kind  at maximum power was investigated by  \cite{Seifert1,Schmiedl1,Tu,De,Ignacio}, which assumed the time dependence of the effective temperature (e.g., variance of the particle position) during the isothermal process. More recent works \cite{Ignacio1} and \cite{quan} proposed the engineering swift equilibration and the shortcut to isothermality, respectively, which kept the effective temperature during the isothermal process, but provided neither kinetics corresponding to the thermally-adiabatic process nor investigation on the power and efficiency of the heat engine.

On the other hand, independently from the research activities in Brownian heat engines,  Masuda and Nakamura \cite{mas1,mas2,mas3} proposed a way to
accelerate quantum dynamics with use of a characteristic driving
potential determined by the underlying adiabatic wave function.
This kind of acceleration
is called the fast forward, which means to reproduce a series of events or
a history of matters on a shortened time scale, like a rapid
projection of movie films on the screen. 
The fast forward theory constitutes one of the promising ways of shortcuts to adiabaticity (STA) devoted to tailor excitations in nonadiabatic processes\cite{dem1,dem2,ber1,lewi,chen,torr}. This theory revealed the non-equilibrium equation of states for the quantum gas under a rapid piston \cite{Babajanova} and provided a simple protocol to accelerate the adiabatic quantum dynamics of spin clusters\cite{Iwan}. The fast forward theory is also applicable to dynamical construction of classical adiabatic invariant \cite{Jarz}.  It is fascinating to investigate the fast forward of the heat engine which is classical and stochastic, find the fast-forward protocols, and investigate the
power and efficiency of the engine.

In this paper we shall develop the fast-forward theory of the stochastic Carnot-like heat engine driven by a Brownian particle coupled with a time-dependent harmonic potential and working between the high ($T_h$)- and low ($T_c$)-temperature heat reservoirs. The momentum degree of freedom is taken into consideration throughout the paper, since energetic interaction between the particle and heat reservoir is also carried by the momentum exchange between them.
In Section \ref{sec2}, we are concerned with the isothermal process, apply the fast forward theory to Fokker-Planck-Kramers or simply the Kramers equation, obtain the fast-forward protocols, and calculate both the reversible and irreversible works. In Section \ref{sec3}, we treat the thermally-adiabatic process where there is no averaged heat transfer between the system and heat reservoir, find a fast-forward protocol
which shows a crucial role of momentum degree of freedom, and obtain the reversible work. In Section \ref{sec4} the efficiency at maximum power is  calculated and compared with existing references. In Section \ref{sec7}, we describe summary and discussions on general time scaling functions, etc..  Appendices \ref{apdA} and \ref{apdB} are devoted to some theorems associated with the irreversible works during the fast-forward protocols. Appendix \ref{apdC} is the analysis of the overdamped case which shows a problem arising from the thermally-adiabatic process.

Before entering the following Sections, we shall sketch the fast forward(FF) scheme to be used hereafter. From a mathematical view point, FF scheme is a way to solve the inverse problem to find an unknown target partial differential equation(PDE) for a known FF path which is an accelerated variant of the original path (e.g., a parameter-dependent Gaussian probability distribution) constructed from  a solution of the known PDE. In this article, PDE means the Kramers equation.
Then the strategy of FF consists of 2 steps: (i) In the unknown target PDE, Hamiltonian is given by a sum of $H_0(x,p)$ and $h(x,p)$(i.e., an unknown driving protocol), while in the original PDE, Hamiltonian is given only by $H_0(x,p)$. What is nontrivial is the existence of such $h(x,p)$, which is determined by the nature of the non-accelerated original path; (ii) the analysis in the step (i) assumes an extremely slow time evolution of the original path. To see both the FF path and target PDE working on a laboratory time scale, we replace the time variable by its advanced variant generated by a very large FF time scaling factor. Thus we can find the target PDE  which is satisfied by the FF path.
The target PDE obtained in this way includes additional terms related to $h(x,p)$ (see Eq.(\ref{FF-FP})).


\section{Fast-forward of isothermal process}\label{sec2}

\subsection{Derivation of driving potential}
We shall develop the probabilistic theory of the stochastic heat engine using a Brownian particle confined by the harmonic potential which has a time-dependent stiffness coefficient. 


In this Section we develop the fast forward theory for the isothermal process in the Carnot-like cycle. Here the Brownian particle is in touch with a reservoir at temperature $k_B T(=\frac{1}{\beta})$ and working under the expanding or compressing trapping potential. 
In the stochastic energetics \cite{Sekimoto1,Sekimoto2,seki-hond} on which the present article is based, the inertial effect or momentum degree of freedom
plays an essential role. So we shall investigate the underdamped region of a Brownian particle, where the Kramers  equation  
for its distribution function  $\rho_0(x,p,t)$ is derived through the continuity equation \cite{Kampen}
\begin{equation}
\label{cont-eq}
\frac{\partial  \rho_0}{\partial t}+\frac{\partial J_x}{\partial x} +\frac{\partial J_p}{\partial p}=0
\end{equation}
with the probability vector flux $(J_x, J_p)$ defined as
\begin{eqnarray}
\label{flux}
J_x &=&\Bigl(\frac{\partial H_0}{\partial p}+\frac{1}{\beta}\frac{\partial}{\partial p}\Bigr)\rho_0, \nonumber\\  
J_p&=&-\Bigl(\frac{\partial H_0}{\partial x}+\frac{1}{\beta}\frac{\partial}{\partial x}\Bigr)\rho_0 \nonumber\\ 
&-&\gamma\Bigl(\frac{\partial H_0}{\partial p}+\frac{1}{\beta}\frac{\partial}{\partial p}\Bigr)\rho_0.
\end{eqnarray}
Here  $H_0=\frac{p^2}{2}+\frac{1}{2}\lambda x^2$ is Hamiltonian for a particle with unit mass trapped by the harmonic potential with stiffness coefficient
$\lambda$. 
$\gamma$ stands for the friction coefficient responsible to dissipation.  Using Eq.(\ref{flux}), Eq.(\ref{cont-eq}) is rewritten as
\begin{eqnarray}
\frac{\partial \rho_0}{\partial t}&=&\lbrace H_0,\rho_0\rbrace \nonumber\\
&+&\gamma\partial_p( \frac{\partial H_0}{\partial p}\rho_0+\frac{1}{\beta}\partial_p \rho_0), 
\label{origin-FP}
\end{eqnarray}
where  $\lbrace \cdots , \cdots \rbrace$ is the Poisson bracket.
The last term proportional to $\frac{\gamma}{\beta}$  is traced back to the Gaussian white noise in the underlying Langevin equation. 

As for the  the probability vector flux, there are several variants of definition which reproduce Eq.(\ref{origin-FP}).  Among them, however, the definition in Eq.(\ref{flux}) is convenient, because $J_x$ and $J_p$ vanish to observe the detailed balance in equilibrium for a static potential. 

If $\lambda=const.$, we have the equilibrium Gaussian distribution function $\rho_0^{eq}$ at $t \rightarrow \infty$. 
Assuming $\partial_t \rho_0=0$ in Eq.(\ref{origin-FP}),  we see:
\begin{equation}
\label{2.4}
\rho_0^{eq}=\frac{\beta \sqrt{\lambda}}{2 \pi } \exp \left(-\beta H_0(\lambda) \right),
\end{equation}
which fulfills the normalization,
$\int_{-\infty}^{+\infty} \int_{-\infty}^{+\infty} \rho_0^{eq}(x,p) dx dp =1$.

If $\lambda$ will be time dependent, the solution in Eq.(\ref{2.4}) becomes meaningless. But the idea of fast forward can guarantee the form in Eq. (\ref{2.4}), even when $\lambda$ is time dependent. 

The first half of the fast forward scheme is the regularization procedure and the second half is replacement of  the time variable by its future or advanced variant.

Firstly we shall explain the regularization procedure.
Let $\lambda$ vary in time very slowly, namely in a quasi-static way:
\begin{equation}
\label{kappa-eps}
\lambda(t) \equiv  \lambda_0 + \epsilon t
\end{equation}
with the growth rate $|\epsilon| \ll 1$, which means that it requires a very long time $T=O(\frac{1}{|\epsilon|})$, to see a recognizable change of $\lambda(t)$.

We take the regularized distribution function $\rho_0^{reg}(x, p; \lambda(t))$ which has the same functional form as $\rho_0^{eq}$ in Eq.(\ref{2.4}):
\begin{equation}
\label{reg-dis}
\rho_0^{reg}=\exp\Biggl[
-\beta H_0(\lambda(t))-\Gamma(\lambda(t))\Biggr]
\end{equation}
where 
\begin{eqnarray}
H_0(\lambda(t))&\equiv& \frac{p^2}{2}+\frac{\lambda(t)}{2}x^2, \nonumber\\
\exp(-\Gamma(\lambda(t))) &\equiv& \frac{\beta \sqrt{\lambda(t)}}{2\pi}.
\label{Hamil-0}
\end{eqnarray}
Then, adding a potential $\epsilon h$ to $H_0$ in Eq.(\ref{origin-FP}), we regularize the Kramers  equation as
\begin{eqnarray}
\frac{\partial\rho_0^{reg}}{\partial t}&=&\lbrace H_0+\epsilon h,\rho_0^{reg}\rbrace \nonumber\\
&+&\gamma \partial_p( p\rho_0^{reg}+\frac{1}{\beta}\partial_p\rho_0^{reg}) \nonumber\\
&+& \epsilon\gamma\partial_p\bigl(\rho_0^{reg} \frac{\partial h}{\partial p}\bigr).
\label{reg-FP}
\end{eqnarray}
$h=h(x,p; \lambda)$ will be determined so that $\rho_0^{reg}$ in Eq.(\ref{reg-dis}) should satisfy Eq.(\ref{reg-FP}).

Noting
\begin{eqnarray} 
\partial_t\rho_0^{reg}&=&\frac{\partial \rho_0^{reg}}{\partial \lambda}\frac{d \lambda}{dt} \nonumber\\
&=&\epsilon\Biggl[-\frac{\beta}{2}x^2+\frac{1}{2\lambda}\Biggr]\rho_0^{reg},
\end{eqnarray}
let's compare both sides of Eq.(\ref{reg-FP}) in each order of $\epsilon$.  Firstly we obtain the equality of $O$(1): 
\begin{equation}
 \lbrace H_0,\rho_0^{reg}\rbrace
+\gamma \partial_p( p\rho_0^{reg}+\frac{1}{\beta}\partial_p\rho_0^{reg})=0.
\label{O1-reg}
\end{equation}
It is evident that Eq.(\ref{O1-reg}) is satisfied:  By using the expression for $\rho_0^{reg}$ in Eq.(\ref{reg-dis}), 
each of the first and second terms on the left-hand side of  Eq.(\ref{O1-reg}) can be shown to vanish.

Then the equality of $O(\epsilon)$ from Eq.(\ref{reg-FP}) is
\begin{eqnarray}
\Biggl[-\frac{\beta}{2}x^2+\frac{1}{2\lambda}\Biggr]\rho_0^{reg} 
&=&\lbrace h,\rho_0^{reg}\rbrace \nonumber\\
&+&\gamma \partial_p(\rho_0^{reg}\partial_p h),
\label{Oepsl-reg}
\end{eqnarray}
which will determine the function $h$.
Noting that
$\partial_p\rho_0^{reg}=-\beta p\rho_0^{reg}$
and
$\partial_x\rho_0^{reg}=-\beta\lambda x\rho_0^{reg}$, Eq.(\ref{Oepsl-reg})
can be rewritten as
\begin{eqnarray}
-\frac{\beta}{2}x^2+\frac{1}{2\lambda}&=&
+\beta[\lambda
x\partial_p h-p\partial_x h] \nonumber\\
&-&\gamma\beta p\partial_p h+\gamma\partial_{pp} h.
\label{Oepsl-regRV}
\end{eqnarray}

Equation (\ref{Oepsl-regRV}) for $h$ can be solved by assuming
\begin{equation}
h=ap^2+bpx+cx^2.
\label{h-asumo}
\end{equation}
In fact, using Eq.(\ref{h-asumo}) in Eq.(\ref{Oepsl-regRV}) and equating the constant term and each coefficient of $p^2$, $x^2$ and $px$ to be zero, 
we have 4 linear algebraic equations (with rank 3):
\begin{eqnarray}
b +2\gamma a&=&0, \nonumber\\
\lambda b&=&-\frac{1}{2}, \nonumber\\
2\lambda a-2 c-\gamma b&=&0,\nonumber\\
\frac{1}{2\lambda}-2\gamma a &=&0.
\label{abc-sol}
\end{eqnarray}
The solution of Eq.(\ref{abc-sol}) is
$a=\frac{1}{4\gamma \lambda}$, $b=-\frac{1}{2\lambda}$, and $c=\frac{1}{4}(\frac{1}{\gamma}+\frac{\gamma}{\lambda})$.
Hence Eq.(\ref{h-asumo}) reduces to
\begin{equation} 
h=\frac{1}{4\gamma\lambda}p^2-\frac{1}{2\lambda}px+\Bigl(\frac{1}{4\gamma}+\frac{\gamma}{4\lambda}\Bigr)x^2.
\label{h-solut}
\end{equation}
In the above regularization procedure, we suppressed  terms of $\epsilon^2$ and higher orders. This simplification is justified because we shall employ  below
a time scaling factor $\alpha(t)$ of order $\frac{1}{\epsilon}$  so that $\alpha(t)\epsilon$ becomes to be of order of unity.

Next we shall enter  the second half of the fast forward scheme.
The regularized Kramers equation in Eq.(\ref{reg-FP}) and its solution $\rho_0^{reg}$ in Eq.(\ref{reg-dis}) work well for a long time but on an extremely slow time scale, which is not convenient for experimentalists who want to see the time evolution of the distribution function on a laboratory time scale. This problem can be resolved by replacing the time variable ($t$) appearing in Eqs.(\ref{reg-dis}) and (\ref{reg-FP})  by its advanced or future variant ($\Lambda(t)$) generated by a very large fast-forward time scaling factor $\alpha(t) \gg 1$ \cite{mas2} as
\begin{equation}
\Lambda(t)=\int_0^{t}\alpha(t') dt'.
\label{future}
\end{equation}

In the fast-forward(FF) range ($0\leq t \leq T_{FF} $), $\alpha(t)$ is written as
$\alpha(t)=1+ (\bar{\alpha}-1)f(s)$ with $s \equiv \frac{t}{T_{FF}}$. 
$\bar{\alpha}(>1)$ is the mean value of $\alpha(t)$ and is given by $\bar{\alpha} = T/T_{FF}$. 
Here $T=O(\frac{1}{|\epsilon|})$ is a long time interval of the quasi-static isothermal process and $T_{FF}$
is an arbitrarily short time to reproduce this process. 
 $f(s)(\ge 0)$ is assumed to satisfy
 the boundary condition
$f(0)=f(1)=\dot{f}(0)=\dot{f}(1)=0$ and $\bar{f}= \int_0^1f(s')ds'=1$, and is symmetric w.r.t. $s=\frac{1}{2}$ (i.e., the center of FF range $0\le s \le 1$). 
Among a wide family of functions of $f(s)$, we choose the simplest function $1-\cos(2\pi s)$. Then
\begin{equation}
\alpha(t) = \bar{\alpha}-(\bar{\alpha}-1) \cos(\frac{2 \pi}{T_{FF}}t).
\label{alp-scal}
\end{equation}
While all the results hereafter will depend on this specific choice, our conclusion about the efficiency of the heat engine at its maximum power 
will be universal and  not be affected by the choice of $f(s)$, which we shall elucidate in Section \ref{sec7}.

After the above time scaling, the stiffness coefficient $\lambda$
now varies in time rapidly as
\begin{equation}
\lambda( \Lambda(t))=\lambda_0+\epsilon \Lambda(t).
\end{equation}

Let's define the fast-forwarded distribution $\rho_{FF}$ as 
\begin{eqnarray}
\rho_{FF}(x,p,t) &\equiv & \rho_0^{reg}(x,p;\lambda(\Lambda(t))) \nonumber\\
&=&\exp\Biggl[
-\beta H_0(\lambda(\Lambda(t)))-\Gamma(\lambda(\Lambda(t)))\Biggr].\nonumber\\
\label{2.20}
\end{eqnarray}

Then $\rho_{FF}$ satisfies the same Kramers equation as Eq.(\ref{reg-FP}) with $\epsilon$ prior to $h$ and $\lambda(t)$ included in $H_0, h$ being replaced by 
$\epsilon \alpha$ and  $\lambda(\Lambda(t))$, respectively.
In fact, the time derivative of $\rho_{FF}$ becomes
\begin{eqnarray}
\frac{\partial \rho_{FF}}{\partial t}&=&\frac{\partial \lambda}{\partial t} \frac{\partial \rho_{FF}}{\partial \lambda} 
=\epsilon \alpha \frac{\partial \rho_{FF}}{\partial \lambda} \nonumber\\
&=&\epsilon\alpha\lbrace h,\rho_0^{reg}\rbrace + \gamma \epsilon\alpha\partial_p(\rho_0^{reg}\partial_p h),\nonumber\\
\label{Lamb}
\end{eqnarray}
where the 3rd equality comes from the fast-forward variant of Eq.(\ref{Oepsl-reg}).
The remaining terms on the right-hand side of Eq.(\ref{reg-FP}) proves vanishing, which is the fast-forward version ($t \rightarrow \Lambda(t)$) of 
Eq.(\ref{O1-reg}).

Taking the asymptotic limit
$ \lim_{\epsilon\to 0, \bar{\alpha} \to \infty} \epsilon \bar{\alpha} = \bar{v}$ with $\bar{v}>0$ ($\bar{v}<0$) for $\epsilon\to+0$ ($\epsilon\to-0$) ,
we obtain the
Kramers equation working for the rapid-time scale region:
\begin{eqnarray}
\frac{\partial\rho_{FF}}{\partial t}&=&\lbrace H_0+v(t)h,\rho_{FF}\rbrace \nonumber\\
&+&\gamma\partial_p( p\rho_{FF}+\frac{1}{\beta}\partial_p\rho_{FF}) \nonumber\\
&+& \gamma\partial_p\bigl(\rho_{FF} \frac{\partial (v(t)h)}{\partial p}\bigr).
\label{FF-FP}
\end{eqnarray}

Here $v(t)$ is a velocity function available from $\alpha(t)$ in the asymptotic limit \cite{mas2}:
\begin{eqnarray}
v(t) = \lim_{\epsilon\to 0, \bar{\alpha} \to \infty} \epsilon \alpha(t) = \bar{v}\left(1-\cos\left( \frac{2 \pi}{T_{FF}}t\right)\right).
\label{2.22}
\end{eqnarray}
Consequently, for $0  \le t \le T_{FF}$,
\begin{eqnarray}
\lambda(\Lambda(t))&=&\lambda_0+\lim_{\epsilon\rightarrow 0, \bar{\alpha}\rightarrow \infty}\epsilon\Lambda(t) 
=\lambda_{0}+\int^{t}_{0}v(t')dt'\nonumber\\
&=& \lambda_{0}+\bar{v}T_{FF}\left[\frac{t}{T_{FF}}-\frac{1}{2\pi}\sin\left(\frac{2\pi}{T_{FF}}t\right)\right]. \nonumber\\
\label{reduce-lamda}
\end{eqnarray}

From now on we take the following prescription:
\begin{eqnarray}
\lambda(t) &\equiv &\lambda(\Lambda(t)),\nonumber\\
 \dot{\lambda} (\equiv  \frac{d \lambda}{dt} )& \equiv &\frac{d \lambda(\Lambda(t))}{dt}=v(t) .
\label{new-kap}
\end{eqnarray}
Then we see the values of $\lambda$ at the initial and final stages of the FF dynamics:
\begin{equation}
\lambda (0)=\lambda_0, \quad \lambda (T_{FF})=\lambda_0+\bar{v}T_{FF}
\label{bound}
\end{equation}
and
\begin{equation}
\dot{\lambda}(0)=\dot{\lambda}(T_{FF})=0.
\label{dotbound}
\end{equation}

\begin{figure}[h]
\includegraphics[scale=0.6]{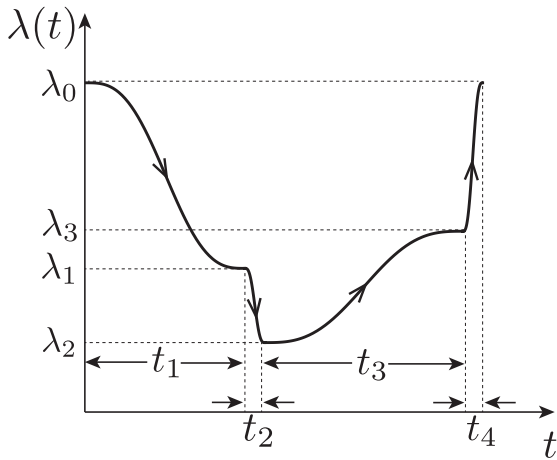}
\caption{Schematic curves for stiffness coefficient  ($\lambda$) as a function of  time ($t$) in the Carnot-like cycle. $t_1,t_2,t_3$ and $t_4$ are the fast-forward time ($T_{FF}$) in each of sub-processes.}
\label{lamb-t}
\end{figure}

Figure \ref{lamb-t} shows schematic curves of $\lambda(t)$ in each of fast-forwarded isothermal and adiabatic processes. 
The curve from $\lambda_0$($\lambda_2$) to $\lambda_1$($\lambda_3$)
with time interval $T_{FF}=t_1$ ($T_{FF}=t_3$) corresponds to isothermal expansion (compression) of the system in contact with the high- (low-) temperature reservoir.

Now we have obtained the fast forwarded Hamiltonian for the particle as:
\begin{equation}
\label{2.29}
H_{FF}(x,p,t)=H_0+\dot{\lambda}h,
\end{equation}
where $H_0$ and $h$  are defined respectively in Eqs.(\ref{Hamil-0}) and (\ref{h-solut}), with $\lambda$ being replaced by its FF version 
in Eq. (\ref{reduce-lamda}).  $H_{FF}(x,p,t)$ is varied during the time interval $0 \le t \le T_{FF}$ at a constant temperature $k_BT=\frac{1}{\beta}$. 

\subsection{Work and heat}
With help of the FF Hamiltonian in Eq.(\ref{2.29})  and the FF Gaussian distribution $\rho_{FF}$ in Eq.(\ref{2.20}), we shall evaluate the thermodynamic quantities, i.e., work $W$, heat $Q$, and internal energy $E$. 
The mean work $W$ done from outside is
\begin{equation}
W=\int\limits_{0}^{T_{FF}}dt\langle\frac{\partial H_{FF}}{\partial t}\rangle.
\label{iso-W-tot}
\end{equation}
Noting
\begin{eqnarray}
\langle\frac{\partial H_{FF}}{\partial t}\rangle&=&\int_{-\infty}^{\infty}\int_{-\infty}^{\infty}\frac{\partial H_{FF}}{\partial t}\rho_{FF}dx dp \nonumber\\
&=&\Biggl(\frac{\dot{\lambda}}{2}-\frac{\gamma\dot{\lambda^2}}{4\lambda^2}+\ddot{\lambda}\Biggl(\frac{\gamma}{4\lambda}+\frac{1}{4\gamma}\Biggr)\Biggr)\frac{1}{\beta\lambda} \nonumber\\
&+&\Biggl(\frac{\ddot{\lambda}}{4\gamma\lambda}-\frac{\dot{\lambda^2}}{4\gamma\lambda^2}\Biggr)\frac{1}{\beta},
\label{iso-H-tdel}
\end{eqnarray}
the work proves to be a sum of the reversible ($W_{rev}$) and irreversible ($W_{irr}$) parts as:
\begin{equation}
W=W_{rev}+W_{irr}
\label{w-totl}
\end{equation}
with
\begin{equation}
W_{rev}=\frac{1}{2\beta}\ln{\lambda}\Bigg|_{0}^{T_{FF}}=\frac{1}{2\beta}\ln\frac{\lambda(T_{FF})}{\lambda_0}
\label{w-rev}
\end{equation}
and
\begin{equation}
W_{irr}=\Biggl(\frac{\gamma}{8}\int\limits_{0}^{T_{FF}}\frac{\ddot{\lambda}}{\lambda^2}dt+\frac{1}{4\gamma}\int\limits_{0}^{T_{FF}}\frac{\ddot{\lambda}}{\lambda}dt\Biggr)\frac{1}{\beta}.
\label{w-irr}
\end{equation}
In obtaining the compact expression Eq.(\ref{w-irr}) from Eqs.(\ref{iso-W-tot}) and (\ref{iso-H-tdel}),  we used the equality
\begin{equation}
\int\limits_{0}^{T_{FF}}\dot{\lambda}^2\lambda^{-m} dt=\frac{1}{m-1}\int\limits_{0}^{T_{FF}}\ddot{\lambda}\lambda^{-(m-1)}dt 
\label{part-integral}
\end{equation}
with $m > 1$, which can be verified with use of the boundary characteristics in Eq.(\ref{dotbound}). 

The irreversible work $W_{irr}$ in Eq.(\ref{w-irr}) consists of the integral of the type, $\int\limits_{0}^{T_{FF}}\frac{\ddot{\lambda}}{\lambda^n}dt$, which can be expressed in terms of the initial value $\lambda(0)$ and the relative growth rate of $\lambda$ defined by
\begin{eqnarray}
\xi &\equiv& \frac{\lambda(T_{FF})-\lambda(0)}{\lambda(0)}\nonumber \\
&=&\frac{\bar{v}T_{FF}}{\lambda(0)} 
\label{xii}
\end{eqnarray}
during the fast-forwarding time from $t=0$ through $t=T_{FF}$. Using the definition of $\lambda$ in Eqs.(\ref{reduce-lamda}) and (\ref{new-kap}) and making a variable change from
$t$ to $s(\equiv \frac{t}{T_{FF}})$,  we can rewrite  $\int\limits_{0}^{T_{FF}}\frac{\ddot{\lambda}}{\lambda^n}dt$ as
\begin{eqnarray}
\int\limits_{0}^{T_{FF}}\frac{\ddot{\lambda}}{\lambda^n}dt&=&\frac{\bar{v}}{T_{FF}}T_{FF} \nonumber\\
&\times&
\int\limits_{0}^{1}\frac{2\pi \sin(2\pi s)}{\Bigl[\lambda_0+\bar{v}T_{FF}\Bigl(s-\frac{1}{2\pi}\sin(2\pi s)\Bigr)\Bigr]^n} ds \nonumber\\
&=&\frac{1}{T_{FF}}\frac{1}{\lambda_0^{n-1}}Z_n(\xi) ,
\label{formula-Z}
\end{eqnarray}
where
\begin{equation}
Z_n(\xi) \equiv \xi\int\limits_{0}^{1}\frac{2\pi\sin(2\pi s)}{\Bigl[1+\xi\Bigl(s-\frac{1}{2\pi}\sin(2\pi s)\Bigr)\Bigr]^n} ds.
\label{Z-def}
\end{equation}
The expression for $W_{irr}$ is thus given by:
\begin{equation}
W_{irr}=\frac{\gamma k_B T}{8\lambda(0)T_{FF}}Z_2(\xi)+\frac{ k_B T}{4\gamma T_{FF}}Z_1(\xi).
\label{w-irr-isother}
\end{equation}

\begin{figure}[h]
\includegraphics[scale=0.6]{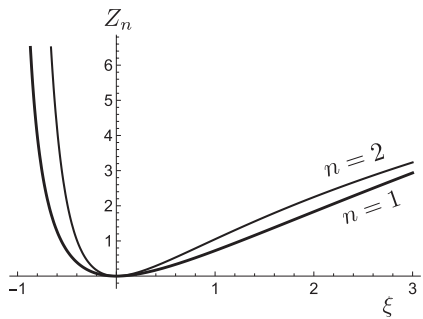}
\caption{Curves of $Z_n(\xi)$ in Eq.(\ref{Z-def}) with $n=1$ and $2$.}
\label{Zn}
\end{figure}

As shown in Fig. \ref{Zn} and in Appendix \ref{apdA}, $Z_1(\xi), Z_2(\xi)$ and thereby $W_{irr}$ are always nonnegative.
The irreversible work  in Eq.(\ref{w-irr-isother}) is inversely proportional to $T_{FF}$, which is consistent with the results of the engineering swift equilibration (overdamped case) \cite{Ignacio1} and of the shortcut to isothermality (underdamped case) \cite{quan}. 
Our new discovery here is that the irreversible work consists of the term proportional to the friction coefficient $\gamma$ and one inversely proportional to
$\gamma$. Note: $\frac{\gamma}{\lambda(0)}$ and $\frac{1}{\gamma}$ has the same dimension under the prescription of unit mass. Alternative derivation of Eq.(\ref{w-irr-isother}) is given in Appendix \ref{apdB}.

In a similar way of calculating the mean work, we obtain the internal energy $E(t)$
\begin{eqnarray}
E(t) &\equiv& \int_{-\infty}^{\infty} \int_{-\infty}^{\infty} H_{FF}\rho_{FF}dx dp \nonumber\\
&=&\Bigl(1+\dot{\lambda}(\frac{1}{2\gamma \lambda}+\frac{\gamma}{4\lambda^2})\Bigr)\frac{1}{\beta}.\nonumber\\
\label{int-E}
\end{eqnarray}
From  Eq.(\ref{int-E}) the increment $\Delta E$ during the fast forward of the isothermal process is found to be given by
$\Delta E=E(T_{FF})-E(0)$=0 because of the boundary values of $\dot{\lambda}$ in Eq.(\ref{dotbound}).

In the isothermal process, the first law of thermodynamics
\begin{equation}
\Delta E =Q+W
\end{equation}
together with $\Delta E=0$ determines the heat $Q$ taken from the heat bath at temperature $T$ as
\begin{equation}
\label{2.58}
Q=-W=-(W_{rev}+W_{irr}).
\end{equation}

\section{Fast forward of thermally-adiabatic process}\label{sec3}
In this Section, we shall  embark upon the fast forward of  the thermally-adiabatic process, where the particle is isolated from a reservoir and working under the expanding or compressing trapping potential. 
In the case of the stochastic microscopic heat engine, however, the unambiguous treatment of the thermally-adiabatic process is controversial and has not yet been settled \cite{seki-hond,Tu,bo,Mart}. 

According to the stochastic energetics, both frictions and random force contribute to the heat transfer between the system and its surrounding\cite{Sekimoto1,Sekimoto2,seki-hond}. 
Since coupling-uncoupling of the particle  system from the environment  is hard, we want to implement thermally-adiabatic process while the particle system is coupled to the environment  with gradually increasing or decreasing temperature. 
Consequently we shall choose a strategy of using the Gaussian distribution function and Kramers equation with the inverse temperature ($\beta$) changing smoothly in a way that guarantees the vanishing heat transfer between a system and the reservoir.  Here we must investigate the mean adiabatic process, since the  Liouville theorem for the individual system does not hold and each system can exchange heat with the bath.

\subsection{Derivation of driving potential}
We shall again apply the fast-forward scheme which consists of the regularization of Kramers equation and the fast forward time-rescaling. The regularized Kramers equation takes the same form as Eq.(\ref{reg-FP}), but here the inverse temperature is time dependent through the time-dependent stiffness coefficient $\lambda(t)$, i.e., $\beta=\beta(\lambda(t))$.
The regularized distribution function is defined by
\begin{equation}
\rho_0^{reg}=\exp\Biggl[
-\beta(\lambda(t))H_0(\lambda(t))-\Gamma(\lambda(t))\Biggr],
\label{reg-dist-adi}
\end{equation}
where
$H_0(\lambda(t))$ and $\Gamma(\lambda(t))$ are given below Eq.(\ref{reg-dis}).  The definition of $\lambda(t)$  is traced back to Eq. (\ref{kappa-eps}). 

The left-hand side of Eq.(\ref{reg-FP}) is of $O(\epsilon)$ and is given by
\begin{eqnarray}
\partial_t\rho_0^{reg}&=&\frac{\partial \rho_0^{reg}}{\partial \lambda}\dot{\lambda} \nonumber\\
&=&\epsilon \Biggl[-\frac{\partial\beta}{\partial \lambda}\frac{p^2}{2}-\frac{\beta}{2}x^2
-\frac{\lambda}{2}\frac{\partial\beta}{\partial \lambda}x^2 \nonumber\\
&+&\frac{1}{\beta}\frac{\partial\beta}{\partial\lambda}+\frac{1}{2\lambda}\Biggr]\rho_0^{reg}.
\label{l-epsilon}
\end{eqnarray} 

The right-hand side of Eq.(\ref{reg-FP})  consists of $O(1)$ and $O(\epsilon)$. The
contribution of $O(1)$ vanishes due to Eq.(\ref{O1-reg}) . The contribution of $O(\epsilon)$ is the same as in the isothermal process:
\begin{eqnarray}
\lbrace\epsilon h,\rho_0^{reg}\rbrace &+& \gamma\epsilon\partial_p(\rho_0^{reg}\partial_p h) \nonumber\\
&=&\epsilon\beta[\lambda x\partial_p h-p\partial_x h]\rho_0^{reg} \nonumber \\
&+&\gamma\epsilon [-\beta p\partial_p h+\partial_{pp} h]\rho_0^{reg}. \nonumber\\
\label{r-epsilon}
\end{eqnarray}

We obtain the equation to solve $h$ by equating the right-hand sides of Eqs.(\ref{l-epsilon}) and (\ref{r-epsilon}).  Using the expansion of $h$ as in Eq.(\ref{h-asumo}) and equating the constant term and each coefficient of $p^2$, $x^2$ and $px$ to be zero, we have the following equations.
\begin{eqnarray}
\frac{1}{2}\frac{\partial\beta}{\partial \lambda}&=&\beta b +2\gamma a\beta,\nonumber\\
\beta\lambda b&=&-\frac{\lambda}{2}\frac{\partial\beta}{\partial\lambda }-\frac{\beta}{2}, \nonumber\\
2\lambda\beta a&-&2\beta c-\gamma\beta b=0, \nonumber\\
\frac{1}{\beta}\frac{\partial\beta}{\partial \lambda}&+&\frac{1}{2\lambda}-2\gamma a =0. \nonumber\\
\label{4unknown}
\end{eqnarray}
There are 4  unknowns ($a, b, c$ and $\beta$).  Among 4 equations above, however, the independent ones are 3,
and we need one more independent equation, which will be available by assuming vanishing heat  transfer in the dynamics of regularized equation. 

In  case of  the constant stiffness coefficient $\lambda$, the time derivative of the mean heat absorbed from the reservoir is:
\begin{eqnarray}
\frac{d Q}{dt}=\int_{-\infty}^{\infty} \int_{-\infty}^{\infty} dx dp\Bigl(J_x\frac{\partial H_0}{\partial x}+J_p\frac{\partial H_0}{\partial p}\Bigr)
\label{def-heat}
\end{eqnarray}
with the probability vector flux $(J_x, J_p)$ in Eq.(\ref{flux}).

When $\lambda$ changes very slowly in time as in Eq.(\ref{kappa-eps}), the regularization procedure in the fast-forward scheme requires
$H_0$ in Eqs.(\ref{def-heat}) and (\ref{flux}) to be replaced by $H_0+\epsilon h(x,p)$, while using the distribution $\rho_0^{reg}$ in
Eq. (\ref{reg-dist-adi}).

On the slow time scale, Eq.(\ref{def-heat}) becomes:
\begin{eqnarray}
\frac{dQ}{dt}&=&\epsilon \int_{-\infty}^{\infty} \int_{-\infty}^{\infty}dx dp
\Bigl(\frac{\partial h}{\partial p}\frac{\partial H_0}{\partial x}\nonumber\\
&-& \frac{\partial h}{\partial x}\frac{\partial H_0}{\partial p}-\gamma \frac{\partial h}{ \partial p}\frac{H_0}{\partial p}\Bigr)\rho_0^{reg}, 
\label{dQdt}
\end{eqnarray}
where $h$ is expanded again as Eq.(\ref{h-asumo}).

Using the expression $\frac{\partial h}{\partial p}=2ap+bx$ and $\frac{\partial h}{\partial x}=bp+2cx$ ,
Eq.(\ref{dQdt}) reduces to
\begin{eqnarray}
\frac{dQ}{dt}=\epsilon (\frac{b}{\beta}-\frac{b+2a\gamma}{\beta})=-\epsilon\frac{2a\gamma}{\beta}
\end{eqnarray}
up to the leading order of $\epsilon$.

Then the vanishing heat transfer ($\frac{dQ}{dt}=0$) during the quasi-static thermally-adiabatic process is satisfied by 
\begin{equation}
a=0.
\label{aaa}
\end{equation}

Using Eq.(\ref{aaa}) in Eq.(\ref{4unknown}), we obtain:
\begin{eqnarray}
b=-\frac{1}{4\lambda},\nonumber\\
c=\frac{\gamma}{8\lambda},\nonumber\\
\beta \sqrt{\lambda}=const.,
\label{sol-therm-ad}
\end{eqnarray}
and consequently the driving potential proves to be
\begin{equation}
h=-\frac{1}{4\lambda}px+\frac{\gamma}{8\lambda}x^2.
\label{adiab-h}
\end{equation}

The second half of the fast-forward scheme in the thermally-adiabatic process is exactly parallel to the description from Eqs.(\ref{future}) through Eq.(\ref{dotbound}) in the isothermal process, except for the difference in the inverse temperature ($\beta$) which is time dependent through $\beta =\frac{const.}{\sqrt{\lambda}}$.

Thus we have the fast-forwarded Hamiltonian $H_{FF}=H_0+\dot{\lambda}h$ with $H_0=\frac{p^2}{2}+\frac{1}{2}\lambda(t) x^2$ and $h$ given in Eq.(\ref{adiab-h}). $\lambda(t)$ is the same as in Eqs.(\ref{reduce-lamda}) and (\ref{new-kap}). The fast-forwarded distribution is $\rho_{FF}(x,p,t)\equiv \rho_0^{reg}(x,p; \lambda(\Lambda(t)))$.

\subsection{Work}
During the fast forward of the thermally-adiabatic process,  the work $W$ done from outside is given by
\begin{equation}
 W=\Delta E=E(T_{FF})-E(0),
 \label{therm-adi-w}
\end{equation}
where $E(t)=\langle H_{FF}\rangle=\int_{-\infty}^{\infty}\int_{-\infty}^{\infty}  H_{FF}\rho_{FF}dx dp$.
Here $H_{FF}=H_0+\dot{\lambda}h$ with $h$ obtained in Eq.(\ref{adiab-h}).
Noting $E(t)=\frac{1}{\beta}+\frac{\gamma \dot{\lambda}}{8\lambda^2}\frac{1}{\beta}$ together with the boundary condition $\dot{\lambda}(T_{FF})=\dot{\lambda}(0)=0$,
we find
\begin{equation}
W=W_{rev}=\frac{1}{\beta(T_{FF})}-\frac{1}{\beta(0)}=k_BT_{final}-k_BT_{initial}.
\label{adia-work}
\end{equation}
In the thermally-adiabatic process we have no irreversible work.

\section{Efficiency at the maximum power of fast-forwarded stochastic heat engine}\label{sec4}

The stochastic engine works between the hot  ($T_h$) and cold ($T_c$) reservoirs.
Using the results of Sections \ref{sec2} and \ref{sec3}, we shall evaluate the efficiency of the fast-forwarded Carnot-like cycle at the maximum power. 
The cycle consists of the following 4 steps as shown in Table 1. See also Fig.\ref{lamb-t}.
We take $\lambda_j$ with $j=0,1,2,3$ as the stiffness coefficients at the nodes of the cycle. 
Among a variety of choices $\{\lambda_j\}$, we concentrate on the symmetric case that the ratio of the initial and final stiffness coefficients is the same in both of the expanding and contracting isothermal processes:
$ \frac{\lambda_1}{\lambda_0}=\frac{\lambda_2}{\lambda_3}=q, $ and
$ \frac{\lambda_2}{\lambda_1}=\frac{\lambda_3}{\lambda_0}=\tilde{q} $.
Table 1 also includes the relative increment of the stiffness coefficient $\xi$,  the time interval $T_{FF}$ (arbitrary) and the mean velocity $\bar{v}$ in each of 4 steps.


\begin{widetext}
{\bf Table 1.} Stiffness coefficients, time interval and mean velocity in each of 4 sub-processes of the accelerated Carnot-like cycle.

\begin{center} 
\begin{tabular}{|l|r|r|r|r|r|r|} \hline  
sub-processes 
& $\lambda(0)$ & $\lambda(T_{FF})$ & $\frac{\lambda(T_{FF})}{\lambda(0)}$ & $\xi \equiv \frac{\lambda(T_{FF})-\lambda(0)}{\lambda(0)}$ 
& $T_{FF}$ & $\bar{v}\equiv  \frac{\lambda(T_{FF})-\lambda(0)}{T_{FF}}$\\ 
\hline  
\hline
1. isothermal expansion at $T_h $& $\lambda_0$ & $\lambda_1$  & $q  (<1)$ & $q-1  (<0)$ &$t_1$ & $\frac{(q-1)\lambda_0}{t_1}$\\
\hline
2. thermally-adiabatic expansion  &
$\lambda_1$ &
$\lambda_2$  &
$\tilde{q} (<1)$ & $\tilde{q}-1 (<0)$ & $t_2$ & $\frac{q(\tilde{q}-1)\lambda_0}{t_2}$\\
\hline
3. isothermal compression at $T_c $&
$\lambda_2$ &
$\lambda_3$  &
$\frac{1}{q}  (>1)$ & $\frac{1}{q}-1  (>0)$ &$t_3$ & $\frac{\tilde{q}(1-q)\lambda_0}{t_3}$\\
\hline
4. thermally-adiabatic compression&
$\lambda_3$ &
$\lambda_0$  &
$\frac{1}{\tilde{q}}  (>1)$ & $\frac{1}{\tilde{q}}-1  (>0)$ & $t_4$ & $\frac{(1-\tilde{q})\lambda_0}{t_4}$\\
\hline

\end{tabular}

\end{center}
\end{widetext}

We now calculate the reversible and irreversible parts of work which the heat engine does on the outside during its one cycle.
In the isothermal processes consisting of the steps 1 and 3,
the reversible part gives a nonvanishing contribution:
\begin{equation}
-\frac{k_B}{2}(T_h-T_c)\ln q,
\end{equation}
while the contribution due to the
irreversible part  is
\begin{eqnarray}
&-&\frac{ k_B}{8}\Biggl(\gamma\frac{Z_{2}(q-1)}{\lambda_0}+\frac{2}{\gamma}Z_{1}(q-1)\Biggr)\frac{T_h}{t_1} \nonumber\\
&-&\frac{ k_B}{8}\Biggl(\gamma\frac{Z_{2}(\frac{1}{q}-1)}{\lambda_2}+\frac{2}{\gamma}Z_{1}(\frac{1}{q}-1)\Biggr)\frac{T_c}{t_3}.\nonumber\\
\end{eqnarray}

In the thermally-adiabatic processes consisting of the steps 2 and 4, we have no irreversible work and the net reversible part gives no contribution:
\begin{equation}
-k_B(T_c-T_h)- k_B(T_h-T_c)=0.
\end{equation}

The total work for one cycle is a sum of contributions from the isothermal and thermally-adiabatic processes and is given by:
\begin{eqnarray}
\label{totl-W}
&&W_{total}=-\frac{k_B}{2}(T_h-T_c)\ln q \nonumber\\
&-&\frac{ k_B}{8}\Biggl(\gamma\frac{Z_{2}(q-1)}{\lambda_0}+\frac{2}{\gamma}Z_{1}(q-1) \Biggr)\frac{T_h}{t_1} \nonumber\\
&-&\frac{ k_B}{8}\Biggl(\gamma\frac{ Z_{2}(q-1)}{\tilde{q}\lambda_0}+\frac{2}{\gamma}Z_{1}(q-1)
 \Biggr)\frac{T_c}{t_3}.\nonumber\\
\end{eqnarray}
On the 3rd line, we
employed the theorem of Appendix \ref{apdA},
\begin{equation}
Z_n(\frac{1}{q}-1)=q^{n-1}Z_n(q-1),
\end{equation}
together with $\lambda_2=\tilde{q}\lambda_1=q\tilde{q}\lambda_0$ available from Table 1.
Concerning the factor $\tilde{q}$, we can see:
\begin{equation}
\tilde{q}=\frac{\lambda_2}{\lambda_1}=\left(\frac{T_c}{T_h}\right)^2
\label{tilde-q}
\end{equation}
with use of the constant of motion in Eq.(\ref{sol-therm-ad}) during the thermally-adiabatic process.

\subsection{Case of large dissipation}
Below we shall first concentrate on the case of a large dissipation with $\gamma \gg \sqrt{\lambda_0}$.
The one-cycle work is expressed as
\begin{eqnarray}
W_{total}&=&-\frac{k_B}{2}(T_h-T_c)\ln q \nonumber\\
&-&\frac{\gamma  k_B}{8\lambda_0}Z_2(q-1)T_h^2\cdot \Biggl(\frac{T_h^{-1}}{t_1}+\frac{T_c^{-1}}{t_3}\Biggr),\nonumber\\
\end{eqnarray}
where the contribution of the term proportional to $\frac{1}{\gamma}$ in Eq.(\ref{totl-W}) is suppressed.
The heat transfer to the particle from the hot heat bath at temperature $T_h$ is given by
\begin{equation}
Q_{in}=-\frac{1}{2} k_B T_h \ln q
-\frac{\gamma  k_B}{8\lambda_0}Z_2(q-1)T_h^2\cdot \frac{T_h^{-1}}{t_1}.
\end{equation}

\begin{figure}[h]
\includegraphics[scale=0.6]{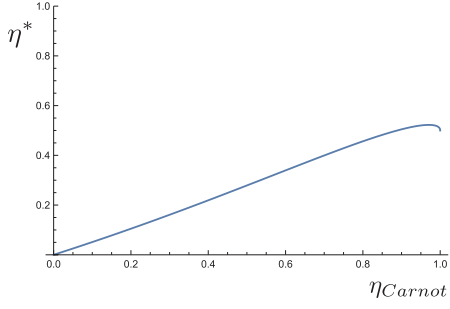}
\caption{$\eta^* $ in Eq.(\ref{eta-stro}) as a function of the Carnot efficiency  $\eta_{Carnot}(\equiv 1-\frac{T_c}{T_h})$.}
\label{fig-rta}
\end{figure}

To obtain a high power heat engine, we take the vanishing time  $t_2=t_4 \rightarrow 0$ with $\bar{v}_2=\bar{v}_4 \rightarrow \infty$ so as to guarantee
$-\frac{\bar{v}_2 t_2}{\lambda_1}=\tilde{q}-1=\left(\frac{T_c}{T_h}\right)^2-1$ and
$\frac{\bar{v}_4 t_4}{\lambda_3}=\frac{1}{\tilde{q}}-1=\left(\frac{T_h}{T_c}\right)^2-1$.
Introducing
\begin{eqnarray}
\label{C.28}
A&=&-\frac{1}{2} k_B \left(T_h-T_c\right)\ln q, \nonumber\\
B&=& \frac{\gamma  k_B}{8\lambda_0}Z_2(q-1)T_h^2
\end{eqnarray}
and assuming $t_2=t_4=0$, the power can be defined by
\begin{equation}
P\equiv \frac{W_{total}}{t_1+t_3}=\frac{A}{t_1+t_3}-\frac{B\left(\frac{T_h^{-1}}{t_1}+\frac{T_c^{-1}}{t_3}\right)}{t_1+t_3}.
\label{PW-stdiss}
\end{equation}
The time
$t_1^*$ and $t_3^*$ which maximizes $P$ is obtained by solving the equations:
\begin{equation}
\frac{\partial P}{\partial t_1}=0, \ \ \frac{\partial P}{\partial t_3}=0,
\end{equation}
which is satisfied by
\begin{equation}
\frac{t_1^*}{t_3^*}=\left(\frac{T_c}{T_h}\right)^{\frac{1}{2}}.
\end{equation}
This issue expresses that $t_1^*$ and $t_3^*$ should be different so as to achieve the maximum power. To be explicit, we have
\begin{eqnarray}
\label{C.37}
t_1^* &=& \frac{2B}{A}\left(T_h^{-1}+(T_h T_c)^{-\frac{1}{2}}\right) , \nonumber\\
t_3^* &=& \frac{2B}{A}\left(T_c^{-1}+(T_h T_c)^{-\frac{1}{2}}\right).
\end{eqnarray}
The efficiency  is written with use of  Eqs. (\ref{C.28})  as
\begin{eqnarray}
\label{eff-bare}
\eta &\equiv& \frac{W_{total}}{Q_{in}} \nonumber\\
&=&\frac{A- B\left(\frac{T_h^{-1}}{t_1}+\frac{T_c^{-1}}{t_3}\right)}{A\frac{T_h}{T_h-T_c}-B\frac{T_h^{-1}}{t_1}}.
\end{eqnarray}
We can express the efficiency at maximum power by substituting Eqs. (\ref{C.37})  into Eq. (\ref{eff-bare}) as
\begin{eqnarray}
\eta^*&=&\frac {1+\sqrt{\frac{T_c}{T_h}}}{2+\sqrt{\frac{T_c}{T_h}}(1+\frac{T_c}{T_h})}\left(1-\frac{T_c}{T_h}\right)\nonumber\\
&=&\frac{1}{2} \left( 1+\frac{1}{2}\left(\frac{T_c}{T_h}\right)^{\frac{1}{2}} - \frac{5}{4}\frac{T_c}{T_h}  
 -\frac{7}{8}\left(\frac{T_c}{T_h}\right)^{\frac{3}{2}} \cdots \right). \nonumber\\
 \label{eta-stro}
\end{eqnarray}
As $\frac{T_c}{T_h}$ increases from zero, $\eta^*$ grows to a maximum and then decreases monotonically.
We see that, for large temperature differences ($T_h \gg T_c$), the limiting efficiency ($\eta^* \rightarrow \frac{1}{2}$) is  
less than that ($\eta^* \rightarrow \frac{2}{3}$) of analysis of the overdamped case \cite{Schmiedl1}.
Figure \ref{fig-rta} shows $\eta^* $ in Eq.(\ref{eta-stro}) as a function of the Carnot efficiency  $\eta_{Carnot}(\equiv 1-\frac{T_c}{T_h})$,
which is available by replacing $\frac{T_c}{T_h}$ in Eq.(\ref{eta-stro}) with $1- \eta_{Carnot}$.

In Appendix \ref{apdC} we showed our application of the FF scheme to the overdamped case.  In the acceleration of the isothermal process, we obtained the results for work and heat analogous to those of the underdamped case, except for the irreversible work $W_{irr}$ which consists of only the term proportional to the friction coefficient in agreement with Ref.\cite{Ignacio1}.
In the acceleration of the thermally-adiabatic process,  however, we encountered  a logical difficulty:
the vanishing of  heat transfer during the thermally-adiabatic process requires
$\frac{\beta}{\lambda}=const.$, which cannot be compatible with real physics where the decrease (increase) of $\lambda$ requires  the cooling (heating) of the system.
Therefore,  the full cycle of the Carnot-like stochastic heat  engine cannot be sketched within a framework of the overdamped case, and it is harmful to conjecture the efficiency of the engine using only the knowledge of the isothermal process  \cite{Schmiedl1}.

\subsection{Case of small dissipation}
Then we consider the case of a small dissipation with $\gamma \ll \sqrt{\lambda_0}$.
The one-cycle work and heat transfer from the hot heat bath are now given by
\begin{eqnarray}
W_{total}&=&-\frac{k_B}{2}(T_h-T_c)\ln q \nonumber\\
&-&\frac{  k_B}{4\gamma}Z_1(q-1)\cdot \Biggl(\frac{T_h}{t_1}+\frac{T_c}{t_3}\Biggr).
\end{eqnarray}
and
\begin{equation}
Q_{in}=-\frac{1}{2} k_B T_h \ln q
-\frac{k_B}{4\gamma}Z_1(q-1)\cdot \frac{T_h}{t_1},
\end{equation}
respectively.

Introducing
$B'= \frac{  k_B}{4\gamma}Z_1(q-1)$ instead of $B$ in Eq.(\ref{C.28})
and assuming $t_2=t_4=0$, the power and efficiency are now given by
\begin{equation}
P\equiv \frac{W_{total}}{t_1+t_3}=\frac{A}{t_1+t_3}-\frac{B'\left(\frac{T_h}{t_1}+\frac{T_c}{t_3}\right)}{t_1+t_3}.
\label{PW-wkdiss}
\end{equation}
and
\begin{eqnarray}
\label{C.39}
\eta &\equiv& \frac{W_{total}}{Q_{in}} \nonumber\\
&=&\frac{A- B'\left(\frac{T_h}{t_1}+\frac{T_c}{t_3}\right)}{A\frac{T_h}{T_h-T_c}-B'\frac{T_h}{t_1}},
\end{eqnarray}
respectively.
The time
$t_1^*$ and $t_3^*$ which maximizes $P$ in Eq.(\ref{PW-wkdiss}) are
\begin{eqnarray}
\label{maxtime}
t_1^* &=& \frac{2B'}{A}\left(T_h+\sqrt{T_h T_c}\right) , \nonumber\\
t_3^* &=& \frac{2B'}{A}\left(T_c+\sqrt{T_h T_c}\right).
\end{eqnarray}
Substituting this $t_1^*$ and $t_3^*$ into Eq.(\ref{C.39}), we have
\begin{equation}
\eta^*= 1-\sqrt{\frac{T_c}{T_h}}. 
\label{CA-issu}
\end{equation}
Interestingly, this result is equal to the Curzon-Ahlborn efficiency for endoreversible heat engines working at maximum power \cite{CA,espo,izumi} although the present stochastic model looks quite different from  macroscopic finite-time heat engines. The issue in Eq.(\ref{CA-issu}) is also compatible with the assertions
of Refs. \cite{Tu,De} which are concerned with the underdamped case, but are solving the equation of motion for  variances of position and momentum of the Brownian particle.

\section{Summary and discussions}\label{sec7}

By extending the idea of the fast forward cultivated in the context of the conservative quantum dynamics, we constructed the fast-forward (FF) theory of the nano-scale stochastic  heat engine driven by a Brownian particle coupled with a time-dependent harmonic potential and working between the high-temperature ($T_h$) and low temperature ($T_c$) heat baths.  The FF scheme applied to the Kramers equation for the underdamped case has successfully reproduced the quasi-static dynamics of the stochastic Carnot-like cycle on the shortened time scale. We have given the explicit expression for the protocols or the driving potentials in both the isothermal and thermally-adiabatic processes, which guarantee the Gaussian probability distribution function throughout the cycle. 
The irreversible work is found to consist of two terms with one proportional to and the other inversely proportional to the friction coefficient. With use of the reversible and irreversible works evaluated by the FF protocols, we have found the efficiency of this engine at maximum power is universal, which is 
$\eta^*=\frac{1}{2} \left( 1+\frac{1}{2}\left(\frac{T_c}{T_h}\right)^{\frac{1}{2}} - \frac{5}{4}\frac{T_c}{T_h}  +O\left(\left(\frac{T_c}{T_h}\right)^{\frac{3}{2}}\right)\right)$   in the case of strong dissipation  and the Curzon-Ahlborn efficiency $\eta^*=1-\left(\frac{T_c}{T_h}\right)^{\frac{1}{2}}$ in the case of weak dissipation.

Our application of FF scheme to the overdamped case in Appendix \ref{apdC} showed that the requirement of the vanishing of heat transfer during the thermally-adiabatic process yields a consequence which is not compatible with real physics. Therefore, the full cycle of the Carnot-like stochastic heat engine can be conceivable only when the momentum degree of freedom is taken into consideration, namely in the underdamped case.

So far we used the time scaling function
$v(t)=\bar{v}(1-\cos(\frac{2\pi}{T_{FF}}t))$ in Eq.(\ref{2.22}) available from $\alpha(t)$ in Eq.(\ref{alp-scal}).
However, it is not necessary to restrict the time scaling function to the above specific function.
Let's define a wide family of the time scaling function
$v(t)=\bar{v}f(s)$ with $s=\frac{t}{T_{FF}}$ in the interval $0 \le s \le 1$, i.e.,   $ 0 \le t \le T_{FF}$. $f(s)(>0)$  is assumed to satisfy the boundary condition
$f(0)=f(1)=\dot{f}(0)=\dot{f}(1)=0$ and $\bar{f}= \int_0^1f(s')ds'=1$, and is symmetric w.r.t. $s=\frac{1}{2}$ (i.e., the center of FF range $0\le s \le 1$). $f(s)$ can include $1-\cos(2\pi s)$, higher harmonics, polynomial functions, etc.
Then the  FF variant of the stiffness coefficient becomes $\lambda=\lambda_0+\int_0^tv(t')dt'=\lambda_0+\bar{v}T_{FF}\int_0^s f(s')ds' \equiv \lambda_0+\bar{v}T_{FF}F(s)$ with $s= \frac{t}{T_{FF}}$,
where $F(s)$ is found to obey $F(0)=0, F(1)=\int_0^1f(s')ds'=1$ and $\dot{F}(0)=\dot{F}(1)=0$.
$F(s)$ includes  the function $s-\frac{1}{2\pi}\sin(2\pi s)$ leading to Eq.(\ref{reduce-lamda}).

With use of the above general scaling function $v(t)$ and $\lambda(t)$, we can reach the same total work as in Eq.(\ref{totl-W}) per cycle of the heat engine,
with $Z_n(\xi)$ defined in a more general form as
$Z_n(\xi)=\xi \int_0^1\frac{\ddot{F}(s)}{\left[1+\xi F(s)\right]^n}ds$. In this general context, Appendix \ref{apdA} already provided the proof for (i) the non-negativity of $Z_n$ and (ii)  the relation $Z_n(\frac{1}{q}-1)=q^{n-1}Z_n(q-1)$, which was essential in our investigation of the power and  efficiency in Section \ref{sec4}. Thus our assertion of the efficiency at the maximum power holds for a wide family of the time scaling functions and the FF protocols to  accelerate the cycle of the heat engine acquire a wide flexibility.

An experimental test of the Brownian heat engine in the underdamped regime can be done with use of levitated cavity opto-mechanics \cite{Ignacio,De2} where the parameters can essentially be tuned independently. The colloidal environment determines the  temperature and Stokes friction for the nanoparticle. 
The harmonic confinement may be realized, via optical tweezers  or via a standing light wave in a cavity.
The external electro-magnetic potential will be useful to mimic the driving protocol which includes the momentum degree of freedom.

{\em Acknowledgments.} We are grateful to Abror Tuymuradov for useful discussions in the early stage of the present work.

\appendix

\section{Proof of  non-negativity of $Z_n(\xi)$ and $Z_n(\frac{1}{q}-1)=q^{n-1}Z_n(q-1)$}\label{apdA}

$Z_n(\xi)$ is defined by
\begin{equation} 
Z_n(\xi)=\xi\int\limits_{0}^{1}\frac{\ddot{F}(s)}{\Bigl[1+\xi F(s)\Bigr]^{n}} ds,
\end{equation}
where $F(s)$ is assumed to be a smoothly growing function of $s$ in the interval $0 \le s \le 1$ and satisfies the boundary condition $F(0)=0$, $F(1)=1$ and  $\dot{F}(0)=\dot{F}(1)=0$.
As described in Section \ref{sec7}, $F(s)$ is constructed from the time-scaling function $f(s)$ that is symmetric w.r.t. $s=\frac{1}{2}$ as
\begin{equation}
\label{f to F}
F(s)=\int_0^s f(s')ds'.
\end{equation}
Although a specific function $F(s)=s-\frac{1}{2\pi}\sin(2\pi s)$ is used in the main text, the proof here is devoted to a broad family of $F(s)$ which includes polynomial functions of $s$.
Firstly we shall show the non-negativity of $Z_n(\xi)$.
Let's  introduce  $g_{\xi}(s)\equiv 1+\xi F(s)$, which is positive for $\xi>-1$ in the interval $0 \le s \le 1$.
Then 
\begin{eqnarray}
Z_n(\xi)&=&\int_0^1 \frac{\ddot{g}_{\xi}(s)}{g_{\xi}^n(s)}ds\nonumber\\
&=&\dot{g}_{\xi}(s)g_{\xi}^{-n}(s)\big|_0^1 +n \int_0^1\dot{g}_{\xi}^2(s) g_{\xi}^{-(n+1)}(s)ds\nonumber\\
&=&n \int_0^1\dot{g}_{\xi}^2 (s)g_{\xi}^{-(n+1)}(s)ds \quad (\ge 0).
\end{eqnarray}
In the last equality, we used $\dot{g}_{\xi}(0)=\dot{g}_{\xi}(1)=1$.
Therefore $Z_n(\xi) \ge 0$ for $\xi >-1$ and the equality holds when $\xi=0$.

Secondly we shall prove the relation $Z_n(\frac{1}{q}-1)=q^{n-1}Z_n(q-1)$.   
$Z_n(\frac{1}{q}-1)$ is explicitly written as
\begin{equation}
Z_n(\frac{1}{q}-1)= \Bigl(\frac{1}{q}-1\Bigr)q^n\int\limits_{0}^{1}\frac{\ddot{F}(s)}{\Bigl[q+(1-q)F(s)\Bigr]^{n}} ds.
\label{Z=prf1}
\end{equation}
If we shall make a variable change
\begin{equation} 
s=1-s',
\end{equation}
then we see the goal:
\begin{eqnarray} 
Z_n(\frac{1}{q}-1)&=& \Bigl(\frac{1}{q}-1\Bigr)q^n \nonumber\\
&\times&\int\limits_{1}^{0}\frac{\ddot{F}(s')ds'}{\Bigl[q+(1-q)\Bigl(1-F(s')\Bigr)\Bigr]^{n}} \nonumber\\
&=&-(1-q)q^{n-1}\nonumber\\
&\times&\int\limits_{0}^{1}\frac{\ddot{F}(s')ds'}{\Bigl[1+(q-1)F(s')\Bigr)\Bigr]^{n}}  \nonumber\\
&=&q^{n-1}Z_n(q-1).
\label{Z=prf2}
\end{eqnarray}
In moving from Eq.(\ref{Z=prf1}) to Eq.(\ref{Z=prf2}), we used  $F(1-s')=1-F(s')$ available from the definition of $F(s)$ in Eq.(\ref{f to F}) applied to the equality for the time-scaling function, $f(s)=f(1-s)$.

\section{Alternative derivation of  $W_{irr}$ in the isothermal process}\label{apdB}
Let us show another derivation of the irreversible work by using the definition,
\begin{equation} 
W_{irr}=T\Delta S-Q,
\end{equation}
where $\Delta S$ and $Q$ are respectively increments of entropy and heat during the isothermal ($\beta=$ constant) process.
We shall extend the definition of $\frac{dQ}{dt}$ in Eq.(\ref{def-heat})  to the fast-forwarded isothermal process,
by replacing $\rho_0$ and $H_0$ by $\rho_{FF}$ in Eq.(\ref{2.20}) and $H_{FF}(=H_0+\dot{\lambda}h)$ in Eq.(\ref{2.29}), respectively. Then
\begin{eqnarray} 
Q&=&\int\limits_{0}^{T_{FF}}dt\int\limits_{-\infty}^{+\infty}dx\int\limits_{-\infty}^{+\infty}dp\Biggl[\dot{\lambda}
\Bigl(\frac{\partial h}{\partial p}\frac{\partial H_0}{\partial x}-\frac{\partial h}{\partial x}\frac{\partial H_0}{\partial p}\nonumber\\
&-&\gamma \frac{\partial h}{\partial p}\frac{\partial H_0}{\partial p}\Bigr)\rho_{FF}
- \dot{\lambda}^2\gamma \Bigl(\frac{\partial h}{\partial p}\Bigr)^2 \rho_{FF}\Biggr]. 
\end{eqnarray}
Noting 
$\frac{\partial h}{\partial p}=\frac{1}{2\gamma\lambda}p-\frac{1}{2\lambda}x$
and
$\frac{\partial h}{\partial x}=-\frac{1}{2\lambda}p+(\frac{1}{2\gamma}+\frac{\gamma}{2\lambda})x$ in the isothermal case,
we obtain:
\begin{equation}
Q=-\int\limits_{0}^{T_{FF}}dt\gamma\Biggr[\frac{\dot{\lambda}}{2\gamma\lambda\beta}+\frac{\dot{\lambda}^2}{4\gamma^2\lambda^2\beta}+
\frac{\dot{\lambda}^2}{4\lambda^3\beta}\Biggl].
\end{equation}

Similarly, with use of the definition of ensemble average of  trajectory entropy \cite{Seifert2, Seifert1,Tu}
\begin{equation}
S \equiv-\int_{-\infty}^{+\infty}dx\int_{-\infty}^{+\infty}dp k_B\rho_{FF}\ln \rho_{FF},
\end{equation}
the increment of the entropy is obtained as:
\begin{eqnarray}
\frac{\Delta S}{k_B}&=&\frac{1}{k_B} \int\limits_{0}^{T_{FF}}\dot{S}dt= \int\limits_{0}^{T_{FF}}dt\int\limits_{-\infty}^{+\infty}dx\int\limits_{-\infty}^{+\infty}dp\gamma \dot{\lambda}\frac{\partial \rho_{FF}}{\partial p}\frac{\partial h}{\partial p} \nonumber\\
&=&\int\limits_{0}^{T_{FF}}dt\int\limits_{-\infty}^{+\infty}dx\int\limits_{-\infty}^{+\infty}dp\gamma\beta\Biggl(-\frac{\dot{\lambda}}{2\gamma\lambda}p^2+\frac{\dot{\lambda}}{2\lambda}px\Biggr)\rho_{FF} \nonumber\\
&=&-\int\limits_{0}^{T_{FF}}dt\frac{\dot{\lambda}}{2\lambda}.
\end{eqnarray}

Then we can evaluate $W_{irr}$ as follows:
\begin{eqnarray} 
W_{irr}&=&-k_BT\int\limits_{0}^{T_{FF}}dt \frac{\dot{\lambda}}{2\lambda} \nonumber\\
&-&\Biggl[-\int\limits_{0}^{T_{FF}}dt\Biggr[\frac{\dot{\lambda}}{2\lambda\beta}+\frac{\dot{\lambda}^2}{4\gamma\lambda^2\beta}+\frac{\gamma\dot{\lambda}^2}{4\lambda^3\beta}\Biggl]\Biggr] \nonumber\\
&=&\int\limits_{0}^{T_{FF}}dt\Biggr[\frac{\dot{\lambda}^2}{4\gamma\lambda^2\beta}+\frac{\gamma\dot{\lambda}^2}{4\lambda^3\beta}\Biggl].
\end{eqnarray}
With use of Eq.(\ref{part-integral}) in the text, the final issue agrees with Eq.(\ref{w-irr}) and thereby leads to Eq.(\ref{w-irr-isother}).

\section{Overdamped case and problem in thermally-adiabatic process}\label{apdC}
Closely following the main text, we apply the FF scheme to the overdamped case of the stochastic heat engine
and show a difficulty encountered in treating the thermally-adiabatic process.

In the isothermal process, the Fokker-Planck equation for the overdamped Brownian particle is given by
\begin{equation}
\label{S2.1}
\partial_t \rho_0(x,t)=-\partial_x j(x,t) 
\end{equation}
with the probability flux
\be
\label{S2.2}
j(x,t) =- \frac{1}{\gamma} \left[\partial_x U_0(x,t)+ \frac{1}{\beta} \partial_x \right]\rho_0(x,t).
\ee
With use of the harmonic potential $U_0(x)=\frac{1}{2}\lambda x^2$, Eq.(\ref{S2.1}) becomes as
\be
\partial_t \rho_0(x,t)=\partial_x\left(\frac{\lambda}{\gamma} x \rho_0\right)+\frac{1}{\beta \gamma} \partial_{xx} \rho_0.
\ee
Assuming $\lambda=const.$, we have the equilibrium distribution at $t \rightarrow \infty$:
\begin{equation}
\label{S2.4}
\rho_0^{eq}(x)=\sqrt{\frac{\lambda \beta}{2 \pi }} \exp \left(-\frac{\lambda \beta}{2}x^2 \right),
\end{equation}
satisfying the normalization $\int_{-\infty}^{+\infty} \rho_0^{eq}(x) dx =1$.

To guarantee the form in Eq. (\ref{S2.4}), even when $\lambda$ is time dependent,
we apply the same FF scheme as in the main text.
Firstly assume $\lambda(t)$ as
\begin{equation}
\lambda \equiv  \lambda_0 + \epsilon t
\end{equation}
with the growth rate $|\epsilon| \ll 1$. We then regularize both the distribution function  and Fokker-Planck equation as
\begin{equation}
\rho_0^{reg}(x,\lambda(t))=\exp \left(-\frac{\beta \lambda(t)}{2}x^2-\Gamma(\lambda(t)) \right)
\end{equation}
with $\exp(-\Gamma(\lambda(t))) \equiv \sqrt{\frac{\beta \lambda(t)}{2\pi}}$
and
\begin{equation}
\partial_t \rho_0^{reg}(x,\lambda(t))=\partial_x \left( \frac{\lambda(t) x+\epsilon \partial_x u}{\gamma} \rho_0^{reg}\right)+\frac{1}{\beta \gamma}\partial_{xx} \rho_0^{reg},
\label{S1}
\end{equation}
where an extra potential $\epsilon u$ is added to $U_0$.
Then the left hand side of Eq.(\ref{S1}) becomes:
\begin{equation}
\label{S5}
\epsilon \frac{\partial}{\partial \lambda}\rho_0^{reg}= \epsilon \left(-\frac{\beta}{2} x^2+\frac{1}{2\lambda}\right) \rho_0^{reg},
\end{equation}
and the right hand side is
\begin{equation}
\label{S6}
\frac{\epsilon}{\gamma}\partial_x(\partial_x u \rho_0^{reg})=\frac{\epsilon}{\gamma}\left((\partial_{xx} u) \rho_0^{reg}+\partial_x u \left(-\beta \lambda x\right) \rho_0^{reg}\right).
\end{equation}
Equating Eq.(\ref{S5}) to Eq.(\ref{S6}), we see the equation for the protocol $u$ as
\begin{equation}
\label{S2.12}
-\frac{\beta}{2}x^2+\frac{1}{2\lambda}=\frac{1}{\gamma}\left(\partial_{xx} u-\beta \lambda x\partial_x u\right),
\end{equation}
which is satisfied by
$u=\frac{\gamma}{4 \lambda}x^2.$

Fast forward version of $\rho_0^{reg}$ is defined by
\begin{eqnarray}
\rho_{FF}&=&\rho_0^{reg}(x,\lambda(\Lambda(t)))\nonumber\\
&=&\exp\left(-\frac{\beta \lambda(\Lambda(t))}{2} x^2-\Gamma(\lambda(\Lambda(t)))\right).
\end{eqnarray}
The  Fokker-Planck  equation working for the rapid-time scale region becomes as $\partial_t \rho_{FF}=\partial_x \left(\frac{\lambda+v(t)\frac{\gamma}{2\lambda}}{\gamma}x \rho_{FF}\right)+\frac{1}{\beta \gamma} \partial_{xx} \rho_{FF}$.
In this way, we see the FF potential
\begin{equation}
U_{FF}=U_0(x)+\dot{\lambda}u(x)=\frac{1}{2}\lambda(\Lambda(t))x^2+\dot{\lambda}\frac{\gamma}{4 \lambda}x^2,
\end{equation}
where the functional $\lambda$ is the same as  in Eqs.(\ref{reduce-lamda}) and (\ref{new-kap}).

The mean work $W$ and heat $Q$ during the isothermal process will be evaluated as below.
The mean work $W$ done from outside is
\be
\label{S2.30}
W= \int_{0}^{T_{FF}} \left\langle \frac{\partial U_{FF}}{\partial t} \right\rangle dt,
\ee
where 
\begin{eqnarray}
\label{S2.31}
\left\langle \frac{\partial U_{FF}}{\partial t} \right\rangle &=&\int_{-\infty}^{+\infty} \rho_{FF}(x,\lambda(\Lambda(t))) \frac{\partial U_{FF}}{\partial t} dx \nonumber\\
&= & \frac{1}{2\beta \lambda} \left(\dot{\lambda}-\frac{\gamma}{2} \frac{\ddot{\lambda} \lambda -\dot{\lambda}^2}{\lambda^2}\right).
\end{eqnarray}

Substituting Eq. (\ref{S2.31}) into Eq. (\ref{S2.30}), we find:
\be
W=W_{rev}+W_{irr}
\ee
with 
\be
W_{rev} =\frac{1}{2\beta}\ln{\lambda}\Bigg|_{0}^{T_{FF}}=\frac{1}{2\beta}\ln\frac{\lambda(T_{FF})}{\lambda_0}
\ee
and
\be
\label{ovd-diiss}
W_{irr}=\frac{ \gamma k_B T}{8\lambda(0)T_{FF}}Z_2(\xi),
\ee
where $\xi$ and $Z_2(\xi)$ are the same as in Eqs.(\ref{xii}) and (\ref{Z-def}), respectively.
Equation (\ref{ovd-diiss}) accords with Ref.\cite{Ignacio1}.
The mean internal energy is
$E(\lambda(\Lambda(t))) \equiv  \int_{-\infty}^{+\infty} dx \rho_{FF} (x, \lambda (\Lambda (t))) U_{FF}(x,\lambda (\Lambda(t)))= \frac{1}{2\beta}\left(1+ \gamma \frac{\dot{\lambda}}{2 \lambda^2 } \right)$.
Noting $\Delta E=0$ and the first law of thermodynamics
$\Delta E =Q+W$, the heat from the reservoir at a fixed temperature is given by $Q= -W$.

On the other hand,   in the thermally-adiabatic process we take the same equation as Eq.(\ref{S2.1}) with Eq.(\ref{S2.2}), and apply the FF scheme.
Here the inverse temperature $\beta$ is assumed to be time dependent as $\beta=\beta(\lambda(t))$.
The regularization procedure is parallel to that of the isothermal case. Equation (\ref{S5}) is now replaced by
\begin{eqnarray}
\label{S7}
\epsilon \frac{\partial}{\partial \lambda}\rho_0^{reg}&=& \epsilon \Bigl(-\frac{\beta}{2} x^2-\frac{\lambda}{2}\frac{\partial \beta}{\partial \lambda}x^2 \nonumber\\
&+&\frac{1}{2}(\frac{1}{\lambda}-\frac{1}{\beta}\frac{\partial \beta}{\partial \lambda})\Bigr) \rho_0^{reg},
\end{eqnarray}
while Eq.(\ref{S6}) remains unchanged. Equating Eq.(\ref{S7}) to Eq.(\ref{S6}), we have the equality:
\begin{eqnarray}
\label{S8}
&&-(\frac{\beta}{2} +\frac{\lambda}{2}\frac{\partial \beta}{\partial \lambda} )x^2\nonumber\\
&+&\frac{1}{2}(\frac{1}{\lambda}-\frac{1}{\beta}\frac{\partial \beta}{\partial \lambda})= \nonumber\\
&=&\frac{1}{\gamma}\left(\partial_{xx} u-\beta \lambda x\partial_x u\right).
\end{eqnarray}
Assuming $u=ax^2$, we have a degenerate equation,
\begin{equation}
\label{S9}
\frac{1}{2\lambda}-\frac{1}{2\beta}\frac{\partial \beta}{\partial \lambda}=\frac{2a}{\gamma}
\end{equation}
for 2 unknowns, $a$ and $\beta$.
One more equation is obtained by investigating the mean heat from the reservoir. In the case of the constant $\lambda$, the time derivative of the mean heat is defined by
\begin{equation}
\frac{d q}{d t}=\int_{-\infty}^{+\infty} dx j(x)\frac{\partial U_0}{\partial x}.
\end{equation}
When $\lambda$ changes in time, the regularization replaces $U_0$ and $\rho_0$ by $U_0+\epsilon u$ and $\rho_0^{reg}$, respectively. Then
\begin{equation}
\frac{d q}{d t}=-\frac{\epsilon}{\gamma} \int_{-\infty}^{+\infty} dx \frac{\partial u}{\partial x}\frac{\partial U_0}{\partial x} \rho_0^{reg}=-\epsilon \frac{2a}{\gamma \beta}.
\end{equation}
The vanishing of  heat transfer during the thermally-adiabatic process requires $a=0$.
Then Eq.(\ref{S9}) gives 
\begin{equation}
\frac{\beta}{\lambda}=const.,
\end{equation}
which cannot be compatible with the physical requirement that in the thermally-adiabatic process, the decrease (increase) of $\lambda$ requires the cooling (heating) of the system.
To conclude this Appendix, the framework of the overdamped case cannot describe the full cycle of the Carnot-like stochastic heat  engine, and it is harmful to conjecture the efficiency of the engine \cite{Schmiedl1}. To resolve this difficulty we must resort to the framework of the underdamped case which includes momentum degree of freedom.

%

\end{document}